# Performance Analysis of Unsupervised LTE Device-to-Device (D2D) Communication

Fabian Eckermann, Julian Freudenthal and Christian Wietfeld
TU Dortmund University, Communication Networks Institute (CNI)
Otto-Hahn-Str. 6, 44227 Dortmund, Germany
E-Mail:{fabian.eckermann, julian.freudenthal, christian.wietfeld}@tu-dortmund.de

*Abstract*—Cellular network technology based device-to-device communication attracts increasing attention for use cases such as the control of autonomous vehicles on the ground and in the air. LTE provides device-to-device communication options, however, the configuration options are manifold (leading to 150+ possible combinations) and therefore the ideal combination of parameters is hard to find. Depending on the use case, either throughput, reliability or latency constraints may be the primary concern of the service provider. In this work we analyze the impact of different configuration settings of unsupervised LTE device-to-device (sidelink) communication on the system performance. Using a simulative approach we vary the length of the PSCCH period and the number of PSCCH subframes and determine the impact of different combinations of those parameters on the resulting latency, reliability and the interarrival times of the received packets. Furthermore we examine the system limitations by a scalability analysis. In this context, we propose a modified HARQ process to mitigate scalability constraints. Our results show that the proposed reduced HARQ retransmission probability can increase the system performance regarding latency and interarrival times as well as the packet transmission reliability for higher channel utilization.

*Index Terms*—MANET, VANET, LTE, Vehicular Communication, Cooperative Communication, Device-to-Device Communication, Mobile Nodes

## I. INTRODUCTION

Device-to-device (D2D) communication in cellular networks was first introduced in Long Term Evolution (LTE) release 12. As D2D communication, often also referred as proximity services (ProSe), enables direct communication between two user equipments (UEs) the term sidelink (opposed to uplink/downlink) was introduced by the 3GPP for the direct transmission. In the remainder of this paper we will use the term sidelink instead of D2D in order to align with the 3GPP terminology.

Sidelink communication in LTE rel. 12 was introduced (among other things) to serve the explicitly expressed interest for an LTE based technology for public-safety-related communication services [1]. The use cases for a direct sidelink communication are manifold: video sharing, gaming, proximity-aware social networking or machine-to-machine communication [2]. In addition sidelink communication can improve spectral efficiency and can increase the total throughput observed in a cell area [3]. In [4] and [5] safety-critical vehicular communication is studied with the outcome of sidelink communication not being suitable for these scenarios. The focus of the aforementioned studies lays on automotive vehicular communication where latency and reliability must be guaranteed at highly scalable scenarios. In [6] an emergency scenario with a system of unmanned aerial vehicles (UAV) in combination with unmanned ground vehicles acting as an autonomous sensor swarm to detect radiological and nuclear material is described. For this and other UAV based scenarios scalability requirements are often lower and an autonomous, decentralized communication like sidelink communication might be favored.

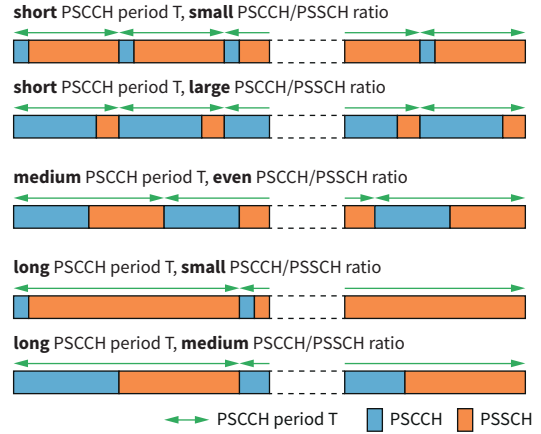

Fig. 1. Exemplary sidelink channel configurations with different PSCCH to PSSCH ratios and PSCCH period lengths.

The various configuration possibilities of sidelink communication, as exemplary depicted by Fig. 1, make it highly adoptable to different scenarios. The performance depends among other things on the PSCCH to PSSCH ratio as well as the number of PSCCH subframes $N_{sf}$ and the period length T, see Section II. Nevertheless tweaking the communication to a desired performance might be complex as the different settings can affect each other and no configuration achieves a global optimum.

The benefits of sidelink communication have been studied by various academic publications [7] [2], however to the best of the authors knowledge as of today no comprehensive analysis of the impact of the configuration settings on the system performance is available. In [8] achievable LTE sidelink performance for platooning with focus on latency is described. The paper uses communication mode 1 (network managed resource allocation, see Section II) and the resource allocation is not analyzed in detail. In [9] improved resource selection approaches for sidelink communication are introduced and compared to LTE rel. 12 sidelink communication in terms of collision probability and throughput. In [10] an analytical model of the resource allocation is introduced and evaluated. A simulation model for the network simulator ns-3 is introduced in [11]. The model is evaluated against analytical results for collision probability and datarate.

The model provided by Rouil et al. [11] is also used as basis for our simulative performance analysis. We investigate the system performance of different sidelink configurations for a scalability analysis and other common evaluation metrics like reliability and latency. In addition for vehicular scenarios frequent update messages of other near by vehicles are important. Therefore the interarrival times of the received packets are evaluated. Based on our results we propose a modified HARQ retransmission scheme to improve the system performance.

We provide the necessary background information on sidelink communication and the modified HARQ scheme in Section II and describe our simulation setup in Section III. Our simulation results are shown in Section IV before the paper is finally concluded in Section V.

## II. SIDELINK COMMUNICATION WITH MODIFIED HARQ RETRANSMISSION SCHEME

### A. Principles of Sidelink Communication

LTE sidelink communication is based on two physical channels [12]:
- Physical sidelink shared channel (PSSCH) that carries the transmission data.
- Physical sidelink control channel (PSCCH) that carries the sidelink control information (SCI) message to detect and decode the PSSCH at the receiving device.

The communication consists of periodically repeating, equally lengths PSCCH periods within each system frame number (SFN) period (1024 frames or 10240 subframes) [1]. Each channel uses a combination of resource blocks (RBs) in the frequency domain and a set of subframes in the time domain. For the resource allocation two modes are defined [12]:
- Mode 1: A specific set of PSCCH/PSSCH resources is assigned to the device by the network. This is only possible for devices under network coverage.
- Mode 2: The device selects the set of PSCCH/PSSCH resources by itself. Mode 2 is independent of the network coverage, however devices without network coverage can of course only operate in Mode 2.

The resource allocation for sidelink communication is shown by Fig. 2. The resources are assigned/selected from a subframe pool that contains all available subframes for the communication. The subframe pool for the PSCCH channel is defined by a PSCCH subframe bitmap and the exact subframes and resource blocks for the transmission are chosen by a parameter delivered by the network (mode 1) or autonomously selected by the transmitting device (mode 2). Each PSCCH transmission is hereby sent twice in different RBs within the same

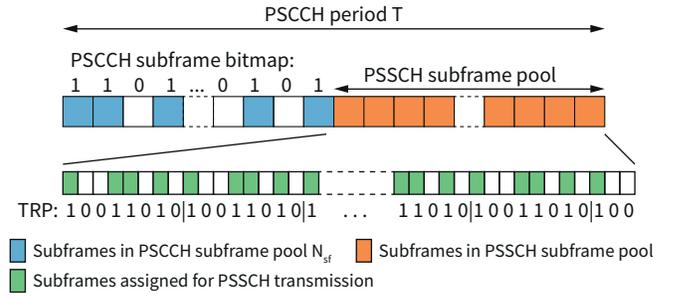

Fig. 2. Resource allocation for sidelink communication in the time domain.

period.

The subframes from the subframe pool for the PSSCH transmission are assigned by a periodic extension of an eight bit Time Repetition Pattern (TRP), which is provided by the network (mode 1) or selected from a TRP table (mode 2) [1]. Furthermore the TRP is part of the SCI message so a receiving device can determine the subframes where the PSSCH is transmitted. The quality of the communication highly depends on the PSCCH to PSSCH ratio. A higher PSCCH to PSSCH ratio adds more configuration overhead to the communication but lowers the probability of PSCCH collisions (see (1) [9] [10]). The collision probability $p_c$ of a PSCCH transmission hereby depends on the number of communication nodes ($N_n$), RBs ($N_{RB}$) and PSCCH subframes ($N_{sf}$).

$$p_c = 1 - (1 - \frac{2}{N_{RB}N_{sf}})^{N_n-1} \quad (1)$$

If both PSCCH transmissions send by a device $N_1$ overlap with a conflicting PSCCH transmission of device $N_2$ the PSSCH transmissions of one or both of the devices may not be received by any device for the whole PSCCH period [9]. The length of the PSCCH period affects the reconfiguration times and therefore how fast PSCCH conflicts are solved. Shorter PSCCH periods add more communication overhead but reduce the propagation of PSCCH conflicts.

For sidelink communication in mode 2 the duty cycle of the PSSCH transmission is limited to maximum 50 %[1]. In Addition every PSSCH transmission is repeated four times due to hybrid automatic repeat requests (HARQ) to overcome interference or collisions with other PSSCH transmissions. So a conflicting PSSCH transmission might be solved by the HARQ process and does not effect the communications performance as much as a PSCCH collision.

### B. Modified HARQ Retransmission Scheme

In addition to a general investigation on the performance of the sidelink communication we study the impact of a modified HARQ retransmission process. As there is no feedback within the HARQ process even successfully transmitted packets will be repeated four times. The retransmissions may collide with other transmissions causing a lower reliability of the communication. To improve this retransmission method for highly loaded communication channels we add a transmission probability to the HARQ process. Every retransmission $A$ is performed with a probability $p_r(A)$.

---
[1]The duty cycle is depicted by the number of ones in the TRP, defining whether or not to use a subframe for transmission. A duty cycle of 50 % corresponds to a TRP with four ones.

## C. Performance Criteria

One key performance parameter is the packet reception ratio (PRR), which is calculated as the number of received packets divided by the number of sent packets. Another parameter is the latency $\tau$ which is determined as time difference between sending and reception of a packet. The last performance criterion is the interarrival time of the received packets (IAT). It is evaluated as the time between the reception of messages from the same transmitter for every transmitter and every receiver. All performance criteria are measured at the application layer.

## III. SIMULATION SETUP

For the simulation we use the network simulator ns-3. The simulation model is based on the LTE D2D model introduced in [11]. All communication nodes are placed such that no packet loss due to channel conditions occurs, the communication channel is therefore not considered within this paper. The devices operate in half-duplex mode so simultaneous transmission and reception of packets is impossible. Unless otherwise noted the simulation parameters are set as listed by TABLE I.
The PSCCH period is set to either 40 ms or 320 ms and the number of PSCCH subframes is set to eight or sixteen to prefer PSSCH over PSCCH transmission. The LTE Modulation and Coding Scheme (MCS) is set to 20 and in order to fit a 100 byte packet into a single subframe four RBs are used per subframe. We deactivated the optional ARQ on the Radio Link Control (RLC) layer to prevent the communication system to transmit outdated data and set the size of the RLC message buffer to match exactly one message in order to minimize the transmission delay. In addition exemplary HARQ probability settings of 0 %, 50 % and 100 % are further investigated.

TABLE I
SIMULATION PARAMETERS

| parameter | value |
| --- | --- |
| simulation time | 60 s |
| transmission interval | 10 ms |
| packetsize | 100 Byte |
| bandwidth | 10 MHz |
| MCS | 20 |
| number of RBs | 4 |
| PSSCH duty cycle | 50 % |

## IV. SIMULATION RESULTS

The impact of a modified HARQ retransmission scheme on the system reliability is shown by Fig. 3. The packet reception ratio for five communication nodes reaches its maximum for the conventional HARQ process (100 % HARQ probability) but it is shown that for a higher number of nodes the packet reception ratio can be increased with a lower HARQ probability. This is further analyzed in the scalability analysis. While the reliability might be decreased the latency is reduced for configurations with decreasing HARQ probability. In Fig. 4 this effect is shown exemplary for a 0 % and 50 % HARQ probability and compared to the conventional process. As defined by the 3GPP a maximum latency of 100 ms for safety-related V2X communication with reliabilities between 80 % and 95 % are required for V2X communication [13]. Our results proof that sidelink communication can only satisfy some of these requirements for a limited number of communication nodes. In a scalability analysis we analyze the system limitations in more detail. In addition to an increasing number of communication nodes we analyzed the outcome of different transmit (tx) intervals and PSSCH duty cycles (see Fig. 5). The results depict that the performance of different sidelink configuration depends on the channel utilization. For a high channel utilization (tx interval = 10 ms) configurations with a small PSSCH subframe pool have a decreased performance. Due to the higher communication overhead less PSSCHs are available and the number of collisions raises. If the ratio of PSCCH and PSSCH subframes is smaller the overhead is decreased. Furthermore a lower PSCCH duty cycle decreases the performance as even for less nodes the channel utilization is to high for the lower number of available PSSCHs. For lower channel utilization (tx interval = 40/50 ms) a lower PSCCH duty cycle increases the performance significantly. The benefit of a synchronization of the transmit interval and the PSCCH period is depicted by the comparison of the transmission intervals of 40 ms and 50 ms. Although the channel utilization of the 40 ms transmit interval is higher the reliability of the system is equal or even increased. If the transmit times of the nodes are not aligned with the PSCCH additional delay is added as the transmission is postponed until resources are assigned. Due to the huge PSSCH pool size the delays exceed the transmission intervals and for the next PSCCH period even more packets must be transmitted so the number of collision increases. For transmit intervals of 50 ms, irrespective of the PSCCH duty cycle, the performance for configurations with more PSSCH subframes is therefore decreased. It is also shown that for all configurations a breakpoint for the number of communication nodes exists where a lower HARQ probability starts to outperform the conventional HARQ procedure.

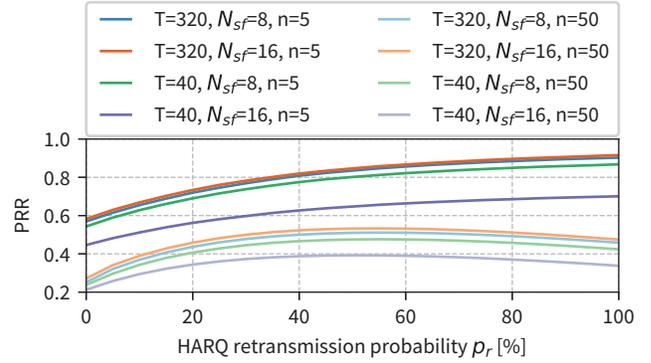

Fig. 3. Impact of different HARQ probabilities on the system reliability (T: PSCCH period [ms], $N_{sf}$: number of PSCCH subframes, n: number of nodes).

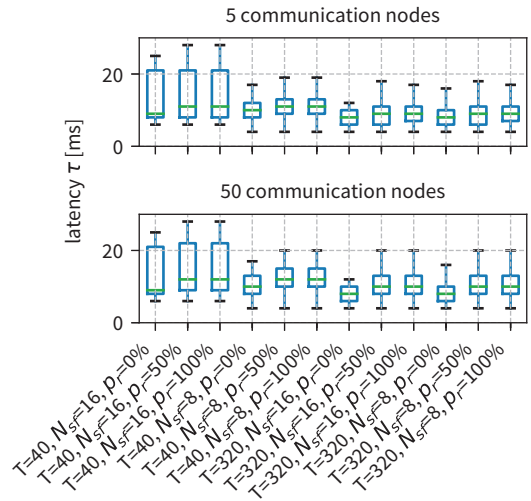

Fig. 4. Latency analysis for different sidelink configurations and HARQ probabilities (T: PSCCH period [ms], $N_{sf}$: number of PSCCH subframes, $p_r$: HARQ probability).

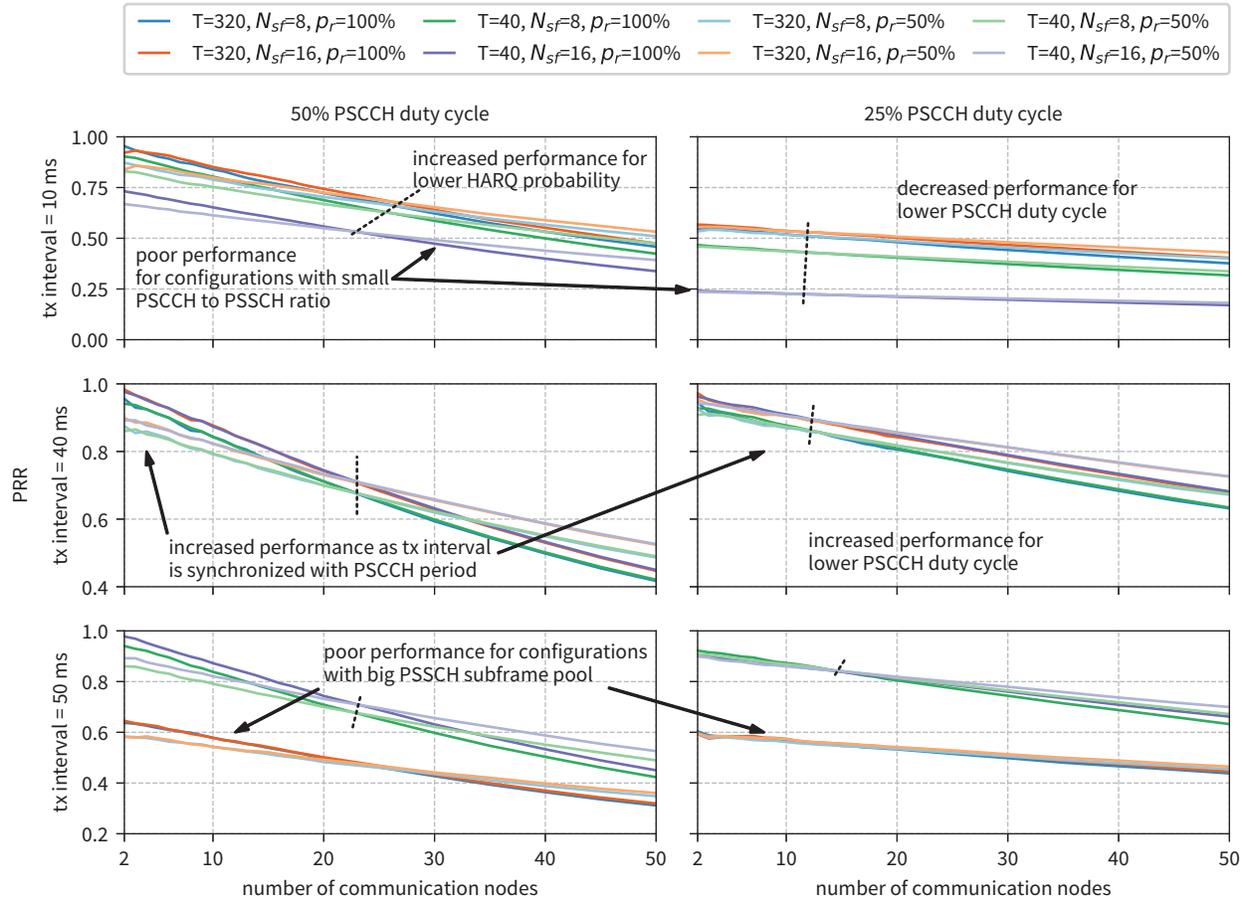

Fig. 5. Scalability analysis for different transmission intervals and PSSCH duty cycles (T: PSCCH period [ms], $N_{sf}$: number of PSCCH subframes, $p_r$: HARQ probability). Note that the dotted line highlights the breakpoint where a lower HARQ probability outperforms the conventional HARQ procedure.

As shown by Fig. 6 the interarrival times for configurations with the maximum PSCCH period (320 ms) are decreased if the HARQ process is completely deactivated. As there is no retransmission procedure every message is either delivered or collides. This results in a lower overall reliability (Fig. 3) but decreases the interarrival time of the delivered messages. This effect is only observed if the number of PSSCH subframes is high. For smaller PSSCH subframe pools the channel utilization is higher so the number of collision increases. A 50 % HARQ probability increases the overall reliability by retransmission of packets which lead to a higher channel utilization. For scenarios with more communication nodes and short PSSCH periods (40 ms) the interarrival times decrease, as due to the retransmissions the collision probability is decreased. For the other configurations the interarrival times ascend. The performance of the conventional HARQ lowers the interarrival times of all configurations except of those with short PSSCH periods and a high number of nodes, compared to the 50 % HARQ probability. For lower PSCCH to PSSCH ratios with less channel utilization the retransmissions can solve conflicts but if the channel utilization becomes overloaded the interarrival times increase.

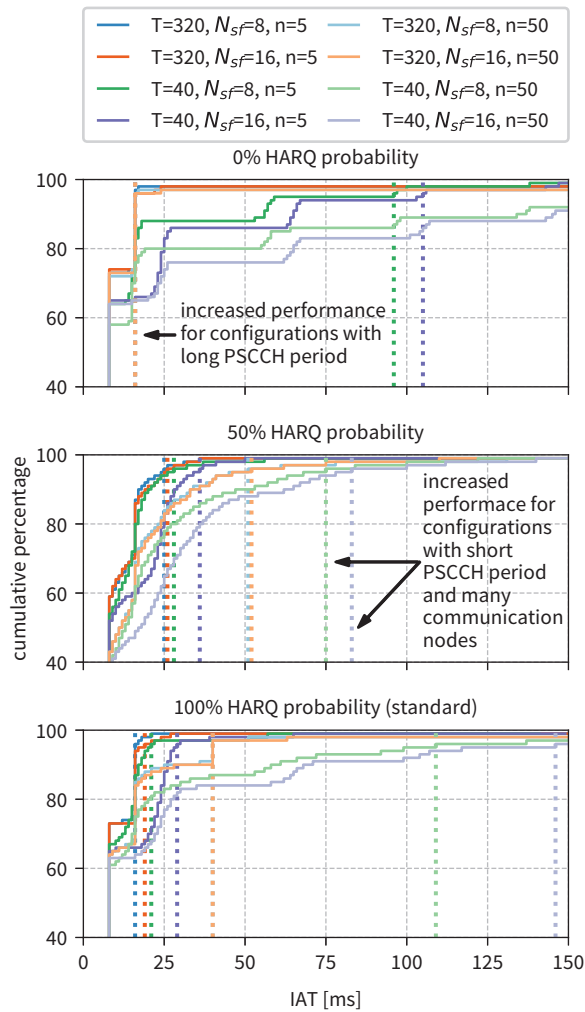

Fig. 6. Impact of the HARQ probability on the interarrival times of the received packets (T: PSCCH period [ms], $N_{sf}$: number of PSCCH subframes, n: number of nodes). Note that the dotted lines mark the 95% quantiles.

It must be noted for the general results, that this is a worst case scenario as there is no randomization of the transmission times of the communication nodes.

## V. CONCLUSION

In this paper we analyzed the system performance for different combinations of PSCCH periods and PSCCH subframe pool sizes. We proposed a probability setting to the HARQ process in order to reduce unnecessary retransmissions. As there is no feedback on the retransmission process this can only be done statistically. We then analyzed the performance and scalability regarding the performance criteria of reliability, latency and interarrival times of the received messages. Our results show that for a low number of communication nodes even the high requirements of safety-critical V2X communication are satisfied. However as the scalability is poor automotive use cases should not focus on sidelink communication. If in contrast the number of nodes is lower (e.g. UAV use cases) or if the latency constrains are less strict (e.g. public safety scenarios) sidelink communication is a valid alternative. In future research we also plan to investigate the performance of mode 4 communication that was added to LTE in rel. 14 specifically addressing vehicular (automotive) communications.

ACKNOWLEDGMENT

The work on this paper has been partially funded by the federal state of Northrhine-Westphalia and the European Regional Development Fund (EFRE) 2014-2020 in the course of the InVerSiV project under grant number EFRE-0800422 and by Deutsche Forschungsgemeinschaft (DFG) within the Collaborative Research Center SFB 876 project B4.

REFERENCES

[1] E. Dahlman, S. Parkvall, and J. Skold, *4G, LTE-Advanced Pro and the Road to 5G (Third Edition)*. Elsevier Science, 2016.
[2] A. Asadi, Q. Wang, and V. Mancuso, "A survey on device-to-device communication in cellular networks," *IEEE Communications Surveys Tutorials*, vol. 16, no. 4, 2014.
[3] K. Doppler, M. Rinne, C. Wijting, C. B. Ribeiro, and K. Hugl, "Device-to-device communication as an underlay to LTE-advanced networks," *IEEE Communications Magazine*, vol. 47, no. 12, December 2009.
[4] L. Gallo and J. Haerri, "Unsupervised long-term evolution device-to-device: A case study for safety-critical V2X communications," *IEEE Vehicular Technology Magazine*, vol. 12, no. 2, June 2017.
[5] R. Molina-Masegosa and J. Gozalvez, "LTE-V for sidelink 5G V2X vehicular communications: A new 5G technology for short-range vehicle-to-everything communications," *IEEE Vehicular Technology Magazine*, vol. 12, no. 4, December 2017.
[6] D. Behnke, S. Rohde, and C. Wietfeld, "Design and experimental validation of UAV-assisted radiological and nuclear sensing," in *15th annual IEEE Symposium on Technologies for Homeland Security (HST 16)*, May 2016.
[7] J. Liu, N. Kato, J. Ma, and N. Kadowaki, "Device-to-device communication in LTE-advanced networks: A survey," *IEEE Communications Surveys Tutorials*, vol. 17, no. 4, 2015.
[8] C. Campolo, A. Molinaro, G. Araniti, and A. O. Berthet, "Better platooning control toward autonomous driving: An LTE device-to-device communications strategy that meets ultralow latency requirements," *IEEE Vehicular Technology Magazine*, vol. 12, no. 1, March 2017.
[9] M. J. Shih, H. H. Liu, W. D. Shen, and H. Y. Wei, "UE autonomous resource selection for D2D communications: Explicit vs. implicit approaches," in *2016 IEEE Conference on Standards for Communications and Networking (CSCN)*, Oct 2016.
[10] D. W. Griffith, F. J. Cintrn, and R. A. Rouil, "Physical sidelink control channel (PSCCH) in mode 2: Performance analysis," in *2017 IEEE International Conference on Communications (ICC)*, May 2017.
[11] R. Rouil, F. J. Cintrón, A. Ben Mosbah, and S. Gamboa, "Implementation and validation of an LTE D2D model for ns-3," in *Proceedings of the Workshop on Ns-3*, 2017.
[12] 3GPP, "Evolved universal terrestrial radio access (E-UTRA); physical layer procedures," 3GPP, TS 36.213 V15.0.0, January 2018.
[13] 3GPP, "Study on LTE support for vehicle-to-everything (V2X) services," 3GPP, TR 22.885 V14.0.0, December 2015.